\documentclass[preprint,12pt]{elsarticle}

\usepackage{amssymb}
\usepackage{amsmath}
\usepackage{hyperref}

\usepackage{comment}
\usepackage{tikz}
\usetikzlibrary{calc}
\usetikzlibrary{shapes,snakes}
\usepackage{subcaption}

\journal{Scientific Computation}

\def\gmn{\Gamma_{m,n}}
\def\gmnk{m_k^2+n_k^2+m_kn_k}
\begin{document}

\begin{frontmatter}



\title{Generating $N$-point spherical configurations with low mesh ratios using spherical area coordinates}


\author{Brian Hamilton}

   \affiliation{organization={Acoustics \& Audio Group, University of Edinburgh},
            addressline={12 Nicolson Sq}, 
            city={Edinburgh},
            postcode={EH8 9DF}, 
            country={UK}}

\begin{abstract}
   This short contribution presents a method for generating $N$-point spherical configurations with low
   mesh ratios.  The method extends Caspar-Klug icosahedral point-grids to
   non-icosahedral nets through the use of planar barycentric coordinates, which are
   subsequently interpreted as spherical area coordinates for spherical point sets.  The
   proposed procedure may be applied iteratively and is parameterised by a sequence of integer
   pairs.  For well-chosen input parameters, the proposed method is able to generate point sets
   with mesh ratios that are lower than previously reported for $N<10^6$.
\end{abstract}

%


\end{frontmatter}


\section{Introduction}
\label{secintro}
Generating a set of $N$ uniformly-distributed points on the unit sphere ($\mathbb{S}^2$) is a
problem with applications in many fields~\cite{saff1997distributing}.  There are
many
approaches to distributing $N$ points on a sphere, including, e.g., Gaussian grids, uniform
random distributions,
spirals~\cite{koay2011analytically},
t-designs~\cite{womersley2018efficient}, and subdivisions of regular
polyhedra~\cite{caspar1962physical,holhocs2014octahedral,michaels2017equidistributed}; 
see~\cite{hardin2016comparison} for a comprehensive review.  There are many measures
to evaluate the quality of such $N$-point spherical configuration, and sequences thereof, including, e.g.,
potential energies, separation and covering distances, and the mesh
ratio~\cite{hardin2016comparison}.  Considering the mesh ratio (where lower is better), it is
known that for a sequence of $N$-point spherical configurations the asymptotic lower bound is
\mbox{$\approx 0.618$}~\cite{bondarenko2014mesh}.  By comparison, it has been shown that a
sequence of equal-area icosahedral points sets has mesh ratios that tend to a value of
\mbox{$\approx 0.697$}, which is the lowest reported thus
far~\cite{hardin2016comparison,michaels2017equidistributed}.

In this short contribution, a method is presented to generate spherical point configurations
with low mesh ratios using Caspar-Klug point sets interpreted as spherical area coordinates
(SAC).  The proposed SAC method is described in detail, subsequently evaluated with examples and
compared to the state of the art, and demonstrates lower mesh ratios than previously reported in
the literature.  Accompanying code for the proposed method is available
at~\url{https://github.com/bsxfun/sac-method}.

\section{Methods}
\subsection{Caspar-Klug point sets}
Caspar and Klug provided a method for generating point sets on an icosahedron based on the
unfolded icosahedral net of triangles superimposed on a triangular grid (hexagonal lattice) of
points~\cite{caspar1962physical}.  This is briefly described below.

Consider the unit vectors $\mathbf{e}_1$ and $\mathbf{e}_2$ in $\mathbb{R}^2$:
\begin{equation}
   \mathbf{e}_1 = (1,0) \,, \quad \mathbf{e}_2 = (1/2,\sqrt{3}/2)
\end{equation}
The following set is a triangular lattice of points generated with the above vectors and the integer pairs $(q_1,q_2)$:
\begin{equation}
   \mathbb{G}=\{\mathbf{r}_{q_1,q_2} = q_1 \mathbf{e}_1 + q_2 \mathbf{e}_2: (q_1,q_2)\in \mathbb{Z}^2\}
\end{equation}
Let $\triangle_{(m,n)}$ denote the triangle with vertices $\mathbf{0}$, $\mathbf{r}_{m,n}$, and
$\mathbf{r}_{-n,m+n}$, and $(m,n)$ is an integer-pair with $m\geq n$, $m>0$, and $n\geq0$.
We define the point set $\mathbb{T}_{m,n}$ as:
\begin{equation}
   \mathbb{T}_{(m,n)}= \{\mathbb{G} \cap \triangle_{(m,n)}\}
\end{equation}
This is illustrated in Figure~\ref{fig1a}.

\newcommand*\rhb{0.45}
\newcommand*\xa{1*\rhb}
\newcommand*\xb{0*\rhb}
\newcommand*\xc{0.5*\rhb}
\newcommand*\xd{0.86603*\rhb}
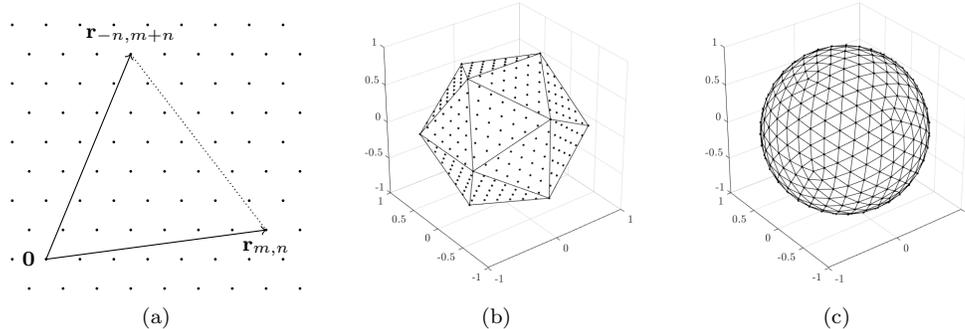
\begin{figure*}[h]
   \begin{subfigure}[b]{0.32\textwidth}
      \centering
      \begin{tikzpicture}
         [hexa/.style= {shape=regular polygon,regular polygon sides=6,minimum
         size=1.16*\rhb cm,inner sep=0,anchor=east,rotate=30,draw=black!50}]
         \foreach \row in {-1,...,7} {
            \foreach \col in {0,2,4,6,8} {
               \fill[black] (\row*\xa,\row*\xb+\col*\xd) circle (0.1ex);
            }
            \foreach \col in {-1,1,3,5,7} {
               \fill[black] (\row*\xa+\xc,\row*\xb+\col*\xd) circle (0.1ex);
            }
         }
         \draw[font=\scriptsize](0,0) node[pos=0.5,left] {$\mathbf{0}$};
         \draw[->,font=\scriptsize] (0,0) -- (6*\xa+1*\xc,6*\xb+1*\xd) node[pos=1.0,below] {$\mathbf{r}_{m,n}$};
         \draw[->,font=\scriptsize] (0,0) -- (-1*\xa+7*\xc,-1*\xb+7*\xd) node[pos=1.0,above] {$\mathbf{r}_{-n,m+n}$};
         \draw[densely dotted,-] (-1*\xa+7*\xc,-1*\xb+7*\xd) -- (6*\xa+1*\xc,6*\xb+1*\xd) node[] {};
      \end{tikzpicture}
      \caption{}
      \label{fig1a}
   \end{subfigure}
   \begin{subfigure}[b]{0.32\textwidth}
      \centering
      \includegraphics[width=\textwidth,clip,trim=2cm 0.5cm 2cm 1cm]{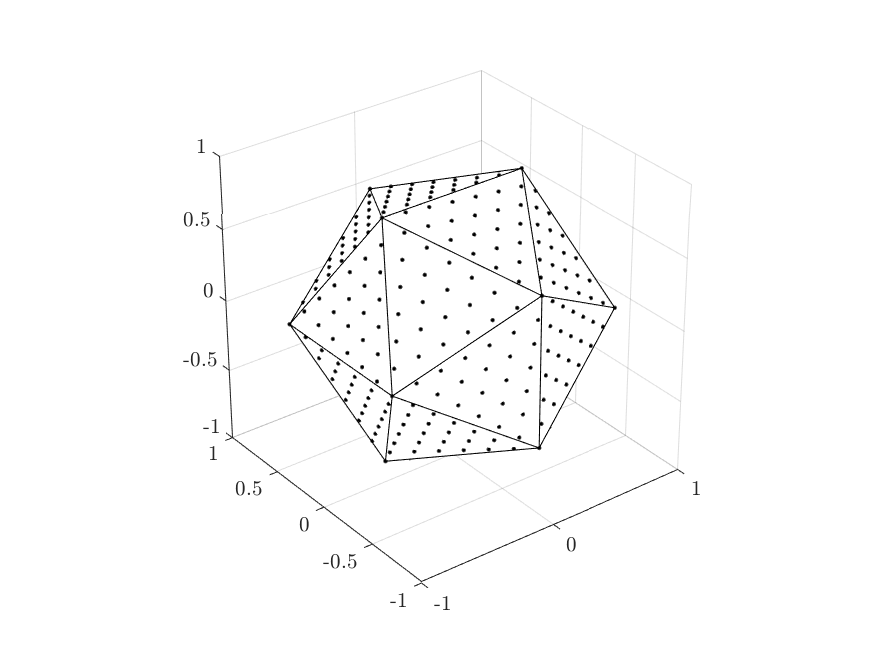}
      \caption{}
      \label{fig1b}
   \end{subfigure}
   \begin{subfigure}[b]{0.32\textwidth}
      \centering
      \includegraphics[width=\textwidth,clip,trim=2cm 0.5cm 2cm 1cm]{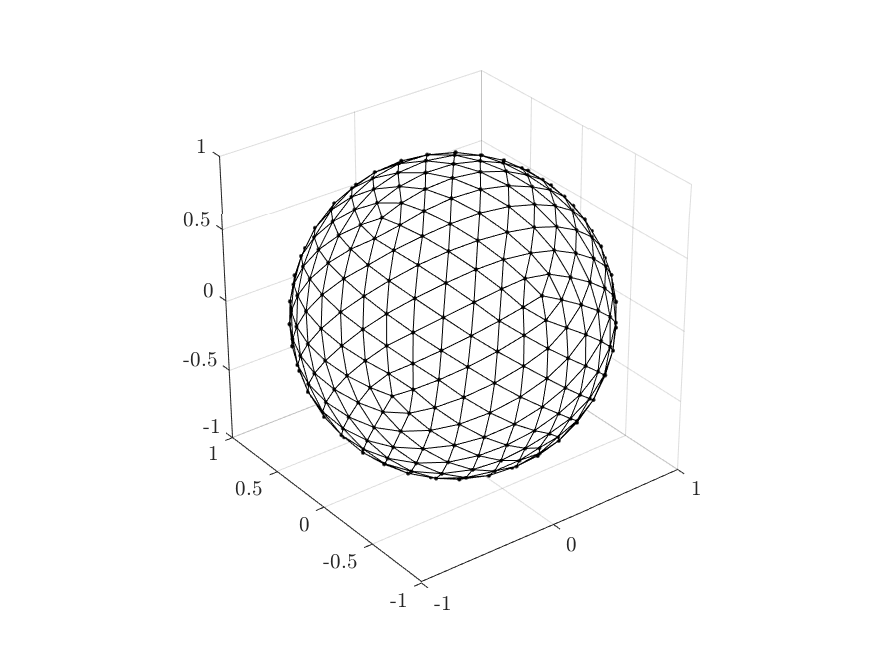}
      \caption{}
      \label{fig1c}
   \end{subfigure}
   \caption{For the choice of $(m,n)=(6,1)$: (a) Illustration of triangular grid, vectors $\mathbf{r}_{m,n}$ and
   $\mathbf{r}_{-n,m+n}$, triangle $\triangle_{(m,n)}$ and point set $\mathbb{T}_{m,n}$; (b) $\mathbb{T}_{m,n}$ on faces of icosahedron; (c)
   projection of icosahedral point set to unit sphere (triangulated with convex hull).}
   \label{fig:hexgrid}
\end{figure*}

Such triangular point grids can be generated over each of the 20 triangular faces of an icosahedron to
give a point set on the surface of the icosahedron~\cite{caspar1962physical} (see Fig.~\ref{fig1b}), and projecting
those points onto the unit sphere gives a spherical point set with icosahedral
symmetry (see Fig.~\ref{fig1c}), also known as ``radial icosahedral
nodes''~\cite{michaels2017equidistributed}.  The cardinality $N$ of such an icosahedral point
set is given by:
\begin{equation}
   N = 10\gmn + 2
\end{equation}
with $\gmn=m^2+n^2+mn$.
This formula was given in~\cite{michaels2017equidistributed} based on the unfolded net of the
icosahedron overlaid on $\mathbb{G}$, and it was also noted that this is independent of the
planar unfolding of the icosahedron, which is ultimately due to the fact that $\triangle_{(m,n)}$ tiles the plane.
As such, this result extends to any unfolded polyhedral net, provided it is made up of 
equilateral triangles.  Thus, the basic operation of generating a point set on an equilateral
triangle can be applied to faces of the regular tetrahedron ($V=4$) or the regular octahedron
($V=6$) to generate spherical point sets with the symmetries of those polyhedra.  Analyzing the
total area of the unfolded net, it can be shown that the cardinality of the resulting point set is
\begin{equation}
   N = (V-2)\gmn + 2
   \label{eqnN}
\end{equation}
with $V\in \{4,6,12\}$.

\subsection{Extension to non-equilateral triangles}
This approach can be extended to the case of non-equilateral triangles by converting the points
in $\mathbb{T}_{m,n}$
to their planar barycentric coordinates.  Let  $\mathbf{a}=\mathbf{r}_{m,n}$,
$\mathbf{b}=\mathbf{r}_{-n,m+n}$, with $m_a=m$, $n_a=n$, $m_b=-n$, and $n_b=m+n$, such that:
\begin{equation}
   \mathbf{a} = m_a \mathbf{e}_1 + n_a \mathbf{e}_2
\,,\quad
   \mathbf{b} = m_b \mathbf{e}_1 + n_b \mathbf{e}_2
\end{equation}
and let $\triangle_{0ab} = \triangle_{(m,n)}$
Next, consider some point $\mathbf{p}=\mathbf{e}_{m_p,n_p} = m_p \mathbf{e}_1 + n_p
\mathbf{e}_1$, such that $\mathbf{p}\in \triangle_{0ab}$ (i.e., $\mathbf{p}\in\mathbb{T}_{m,n}$).
This can be written in terms of three barycentric coordinates $\lambda_{p,a}$, $\lambda_{p,b}$,
and  $\lambda_{p,0}$ as:
\begin{equation}
   \mathbf{p} = \lambda_{p,0} \mathbf{0} + \lambda_{p,a} \mathbf{a} + \lambda_{p,b} \mathbf{b} 
   \label{baryconst1}
\end{equation}
with 
\begin{equation}
   \lambda_{p,a}+\lambda_{p,b}+\lambda_{p,0}=1
   \label{baryconst2}
\end{equation}
and $0\leq \lambda_{p,a},\lambda_{p,b}\leq 1$.  The coordinate $\lambda_{p,0}$ is not needed,
and is only provided for convention.  
Given $(m_p,n_p)$ one can then
calculate $(\lambda_{p,a},\lambda_{p,b})$ from:
\begin{equation}
   \begin{pmatrix}
      \lambda_{p,a} \\
      \lambda_{p,b} 
   \end{pmatrix}= 
   \begin{pmatrix}
      m_a & m_b \\
      n_a & n_b 
   \end{pmatrix}^{-1}
   \begin{pmatrix}
      m_p \\
      n_p
   \end{pmatrix}
=
\frac{1}{\gmn}
   \begin{pmatrix}
      m+n & n \\
      -n & m 
   \end{pmatrix}
   \begin{pmatrix}
      m_p \\
      n_p
   \end{pmatrix}
   \label{bary3}
\end{equation}
Since $\gmn>0$, any valid integer pair $(m_p,n_p)$ corresponding to point $\mathbf{p}$ can
be expressed in planar barycentric coordinates $(\lambda_{p,a},\lambda_{p,b})$ with
respect to $\triangle_{0ab}$.  Having such a grid of barycentric coordinates, one can generate
point sets on any triangle (not necessarily equilateral), simply using~\eqref{baryconst1}.  
Furthermore, one can generate a point set on the surface of any closed
triangular mesh, and if that triangular mesh is the convex hull of its vertices, one can then project
the generated point grid to the unit sphere to obtain a spherical point set.  It follows that
the cardinality of the resulting spherical point set is given by~\eqref{eqnN} for any $V\geq 4$.

Since any such spherical point set can be triangulated via its convex hull, one can also
apply these operations \emph{recursively}.  Starting from a $V_0$-point spherical configuration, recursive point-set generation by the above
procedure results in a spherical point set with cardinality:
\begin{equation}
    N = 2+ (V_0-2)\prod_{k=1}^K(\gmnk)
   \label{eqnNb}
\end{equation}
where $((m_1,n_1),\dots,(m_K,n_K))$ is a sequence of $K>1$ integer-pairs with $m_k\geq
n_k$, $m_k> 0$, and $n_k\geq 0$.

\subsection{Homogenizing with spherical area coordinates}
In order to improve the uniformity of the generated point sets, we can re-interpret the planar
barycentric coordinates as a form of \emph{spherical} barycentric
coordinates~\cite{bauer2000distribution} known as \emph{spherical area
coordinates}~\cite{lei2020new} -- a technique which is known to improve radial icosahedral
point sets (with $K=1$, $n_1=0$ in~\eqref{eqnNb})~\cite{bauer2000distribution,lei2020new}.

For this, consider the planar triangle $\triangle_{0ab}$ with point $\mathbf{p}\in\triangle_{0ab}$.  It is
well-known that for the point $\mathbf{p}$'s planar barycentric coordinates
$\lambda_{p,0},\lambda_{p,a},\lambda_{p,b}$ (given by~\eqref{bary3}  and~\eqref{baryconst2}),
one has the properties that:
\begin{equation}
\frac{|\triangle_{0ap}|}{|\triangle_{0ab}|} = \lambda_{p,b}
\,, 
\frac{|\triangle_{pab}|}{|\triangle_{0ab}|} = \lambda_{p,0}
\,, 
\frac{|\triangle_{0pb}|}{|\triangle_{0ab}|} = \lambda_{p,a}
\label{baryconst3}
\end{equation}
where $|\triangle_{0ab}|$ denotes the planar area of triangle $\triangle_{0ab}$.  

Let $|\triangle_{0ab}|$ now denote the \emph{spherical} area of the
\emph{spherical} triangle $\triangle_{0ab}$.  In the spherical setting, we can use the term ``spherical area
coordinates'' (after~\cite{lei2020new}) to describe
$\lambda_{p,0},\lambda_{p,a},\lambda_{p,b}$  (under the constraints~\eqref{baryconst3}).  However, in this
setting it is well-known that~\eqref{baryconst1} does not hold, and we must instead solve the
system of equations~\eqref{baryconst3} to find
a point $\mathbf{p}$ (given its spherical area coordinates).   Fortunately, an iterative
solution to this problem was provided in~\cite{bauer2000distribution}.  Additionally, an
analytic solution was recently presented in~\cite{lei2020new}.  Formula for that analytic
solution are left out for brevity (see~\cite{lei2020new}), but they are implemented in the
accompanying code.

\section{Evaluations}
Having described the general method for generating spherical point sets from polyhedra with
triangular faces, it remains to evaluate resulting $N$-point spherical configurations.  As a 
quality measure for these point sets we focus on the mesh ratio, which is defined below.  For a $N$-point spherical configuration
$\omega_N$, we first define the separation distance $\delta(\omega_N)$:
\begin{equation}
\delta(\omega_N) = \min_{\substack{x,y\in\omega_N \\ x\ne y}} |x-y|
\end{equation}
and the covering radius $\eta(\omega_N)$:
\begin{equation}
\eta(\omega_N) = \max_{y\in\mathbb{S}^2}\min_{x\in\omega_N} |x-y|
\end{equation}
The mesh ratio is then defined as:
\begin{equation}
\gamma(\omega_N) = \frac{\eta(\omega_N)}{\delta(\omega_N)}
\end{equation}
For this definition of the mesh ratio, it is known that $\lim_{N\to \infty} \gamma(\omega_N)\geq
\frac{\sec(\pi/5)}{2}\approx 0.618$~\cite{bondarenko2014mesh}.

Additionally, for visualization purposes, we consider a triangle quality measure defined here as
the ratio of smallest and largest edge lengths in a triangle (henceforth, ``the edge ratio'').
This edge ratio is non-negative and bounded by one, and the bound one is acheived with an
equilateral triangle (thus, higher is better).  The appearance of higher quality triangles (locally) does not necessarily translate to a
lower mesh ratio (a global measure), but viewing the overall distribution of triangle qualities helps to
provide some insight into the overall mesh ratio.  This triangle quality measure, which is
just one of many possible~\cite{pebay2003analysis}, is chosen for its simplicity to compute.  

For brevity, we consider only icosahedral point sets, wherein the base spherical configuration are $V_0=12$ points
from a regular icosahedron.  Sequences of integer-pairs are herein indicated as
$((m_1,n_1),\dots,(m_K,n_K))$, where a superscript integer indicates repeated entries (e.g.,
$((1,1),(2,0),(2,0))$ can be written $((1,1),(2,0)^2)$.  The SAC method presented here is compared to the
``Equal-area Icosahedral'' (EQA) configurations from~\cite{michaels2017equidistributed}
(computed with the accompanying Matlab code to~\cite{michaels2017equidistributed}).  Note
that the SAC method with $K=1$ and $n_1=0$ is that of~\cite{lei2020new}.

\begin{figure*}[h]
   \begin{subfigure}[b]{0.32\textwidth}
      \centering
      \includegraphics[width=\textwidth,clip,trim=1cm 0.5cm 1cm 0.5cm]{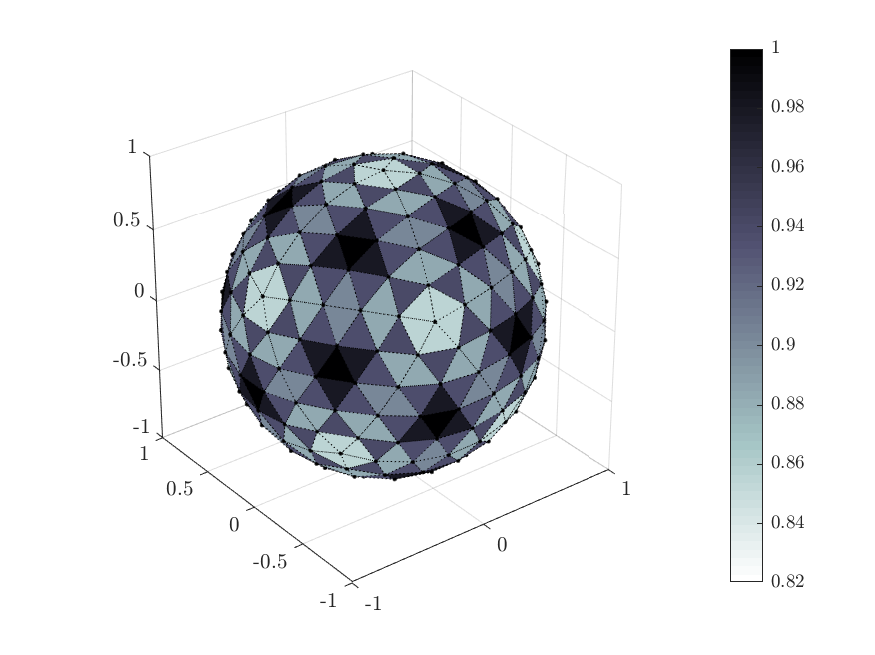}
      \caption{EQA $((5,0))$; $N=252$; $\gamma(\omega_N)=0.663$}
      \label{fig2a}
   \end{subfigure}
   \begin{subfigure}[b]{0.32\textwidth}
      \centering
      \includegraphics[width=\textwidth,clip,trim=1cm 0.5cm 1cm 0.5cm]{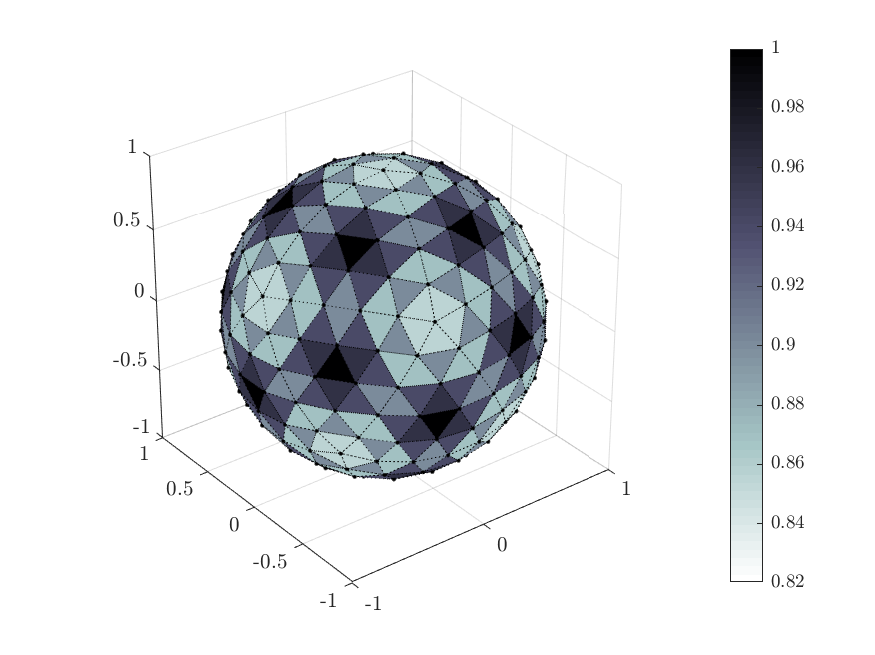}
      \caption{SAC $((5,0))$; $N=252$; $\gamma(\omega_N)=0.653$}
      \label{fig2b}
   \end{subfigure}
   \begin{subfigure}[b]{0.32\textwidth}
      \centering
      \includegraphics[width=\textwidth,clip,trim=1cm 0.5cm 1cm 0.5cm]{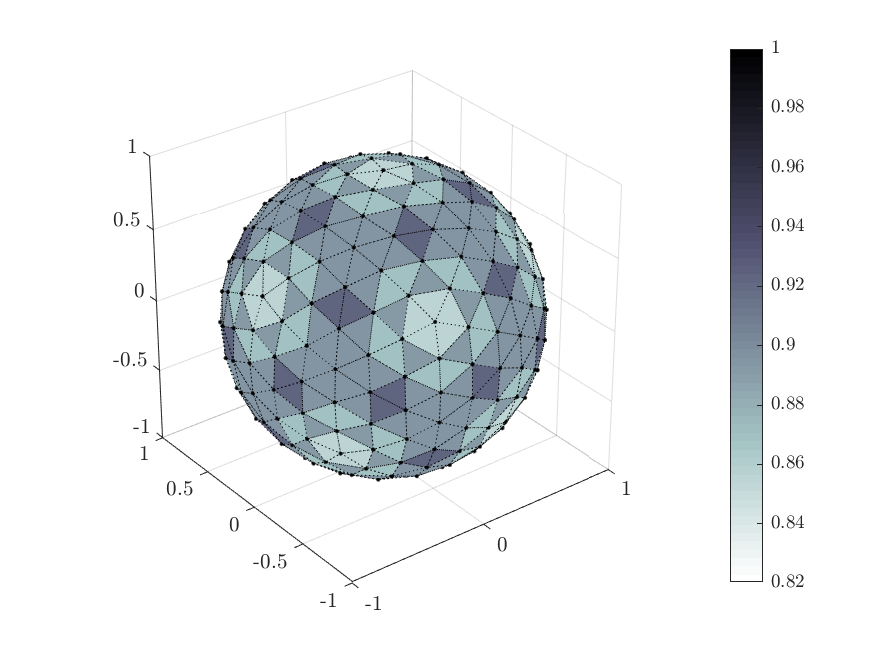}
      \caption{SAC $((1,1),(3,0))$; $N=252$; $\gamma(\omega_N)=0.631$}
      \label{fig2c}
   \end{subfigure}
   \\
   \begin{subfigure}[b]{0.32\textwidth}
      \centering
      \includegraphics[width=\textwidth,clip,trim=1cm 0.5cm 1cm 0.5cm]{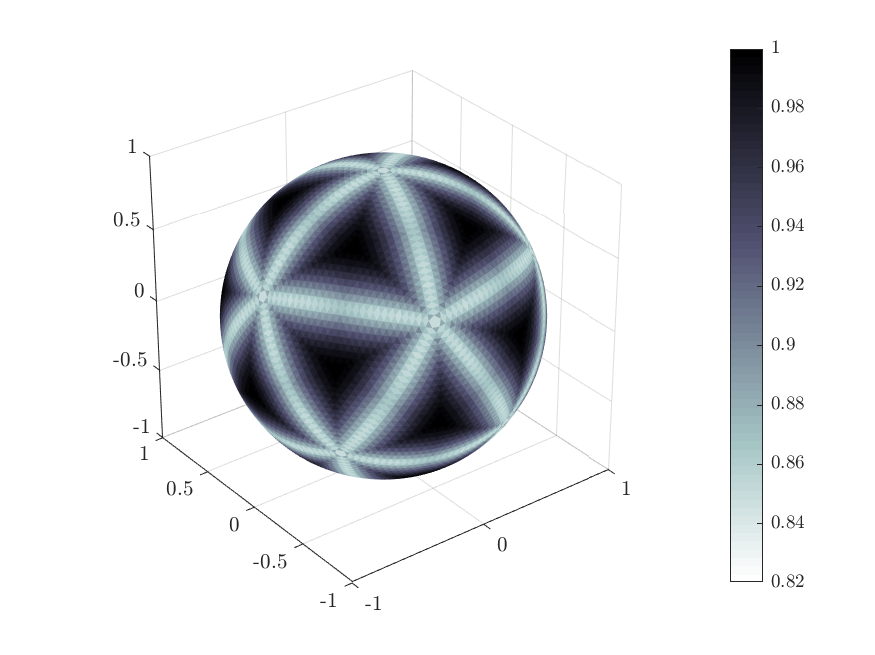}
      \caption{EQA $((27,0))$; $N=7292$; $\gamma(\omega_N)=0.689$}
      \label{fig3a}
   \end{subfigure}
   \begin{subfigure}[b]{0.32\textwidth}
      \centering
      \includegraphics[width=\textwidth,clip,trim=1cm 0.5cm 1cm 0.5cm]{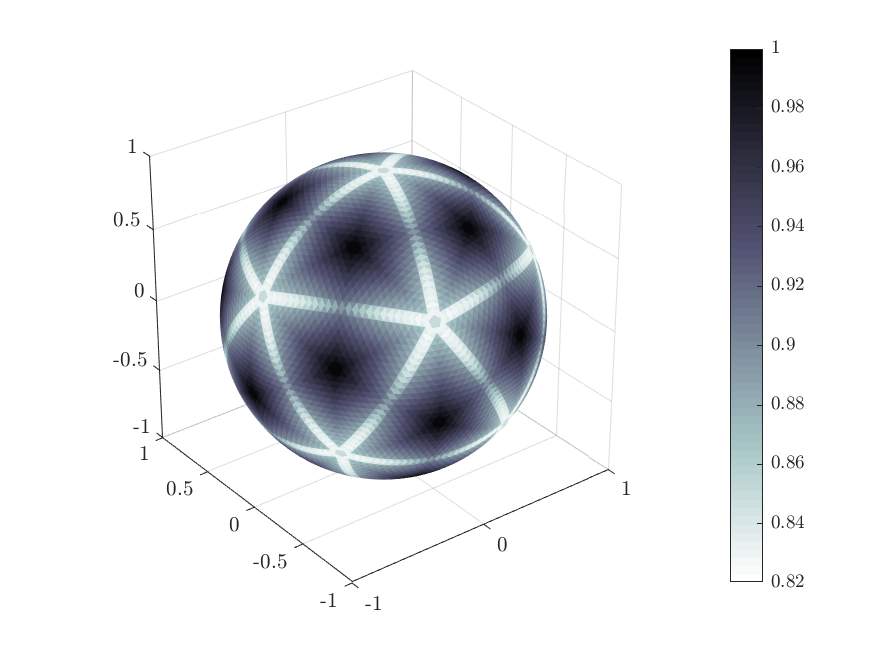}
      \caption{SAC $((27,0))$; $N=7292$; $\gamma(\omega_N)=0.664$}
      \label{fig3b}
   \end{subfigure}
   \begin{subfigure}[b]{0.32\textwidth}
      \centering
      \includegraphics[width=\textwidth,clip,trim=1cm 0.5cm 1cm 0.5cm]{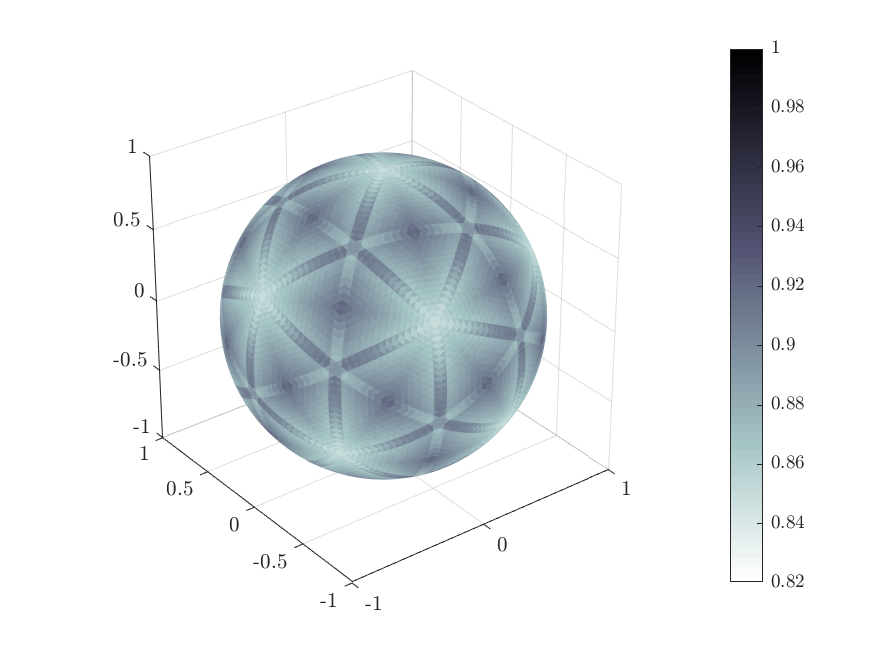}
      \caption{SAC $((1,1),(16,0))$; $N=7682$; $\gamma(\omega_N)=0.644$}
      \label{fig3c}
   \end{subfigure}
   \\
   \begin{subfigure}[b]{0.32\textwidth}
      \centering
      \includegraphics[width=\textwidth,clip,trim=1cm 0.5cm 1cm 0.5cm]{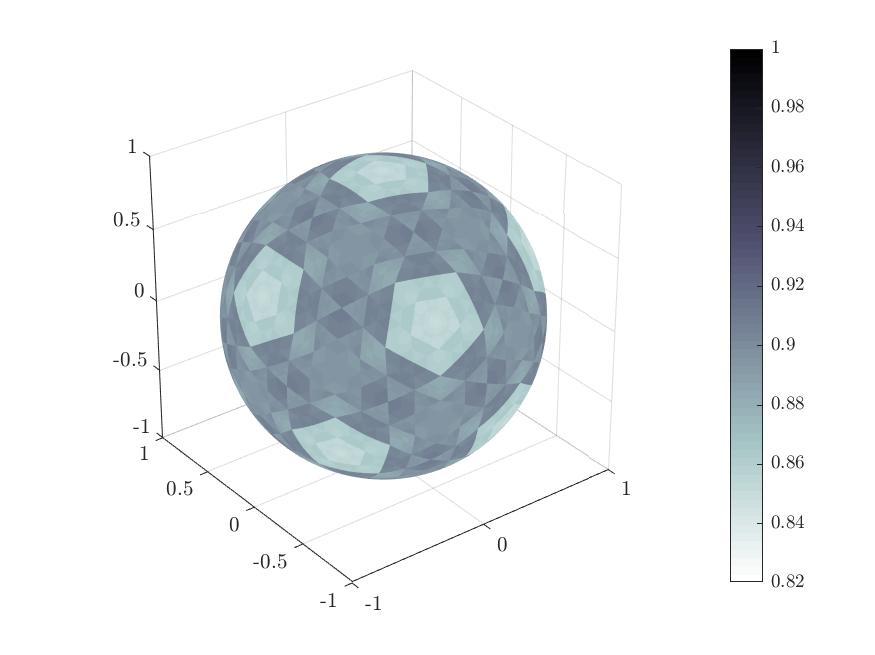}
      \caption{SAC $((1,1),(2,0)^4)$; $N=7682$; $\gamma(\omega_N)=0.639$}
      \label{fig3d}
   \end{subfigure}
   \begin{subfigure}[b]{0.32\textwidth}
      \centering
      \includegraphics[width=\textwidth,clip,trim=1cm 0.5cm 1cm 0.5cm]{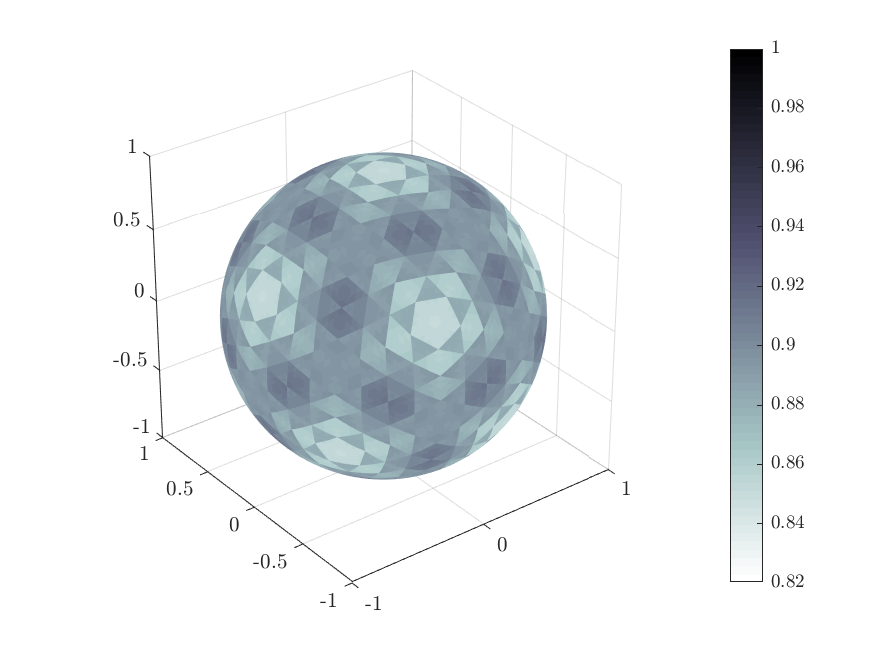}
      \caption{SAC $((1,1),(4,0)^2)$; $N=7682$; $\gamma(\omega_N)=0.630$}
      \label{fig3e}
   \end{subfigure}
   \begin{subfigure}[b]{0.32\textwidth}
      \centering
      \includegraphics[width=\textwidth,clip,trim=1cm 0.5cm 1cm 0.5cm]{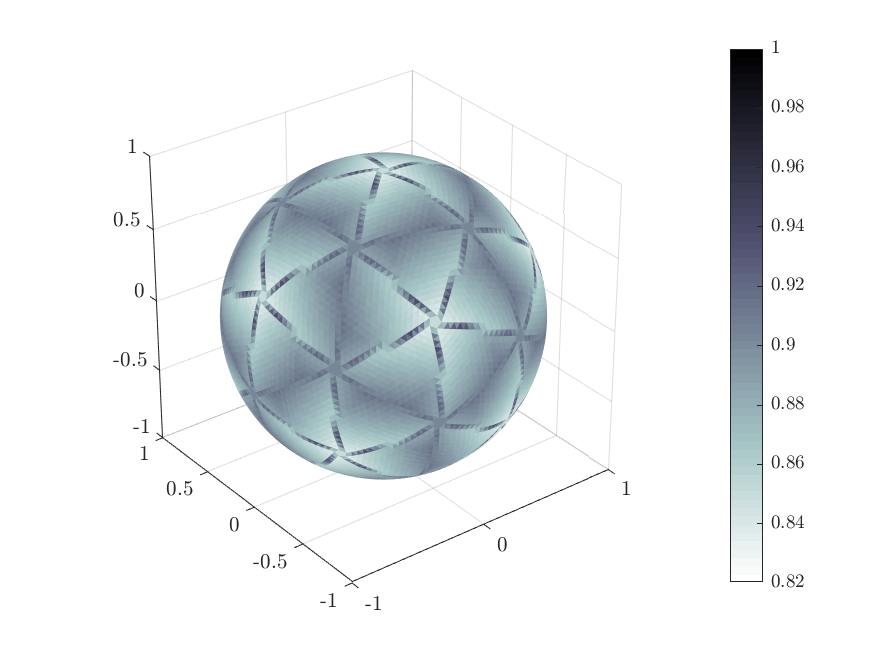}
      \caption{SAC $((1,1),(15,2))$; $N=7772$; $\gamma(\omega_N)=0.643$}
      \label{fig3f}
   \end{subfigure}
   \caption{Illustrations of $N$-point icosahedral spherical configurations with equal-area
   method (EQA) and spherical-area-coordinate (SAC) method for sequence of $(m_k,n_k)$
   integer-pairs.  Calculated mesh ratios are as indicated, and triangles are colored by edge
   ratios.  For ease of viewing, triangle edges and vertices are not displayed for denser point sets
   in Figs.~\ref{fig3a}--\ref{fig3f}.}
   \label{fig3}
\end{figure*}

To start, we consider the configurations with $N=252$ illustrated in Figs.~\ref{fig2a}--\ref{fig2c}.  
ubtle differences between the EQA method
(Fig.~\ref{fig2a}) and the SAC method (Fig.~\ref{fig2b}) may be observed for the integer-pair $(5,0)$.  In particular
the EQA method provides a point set whose triangulation has higher-quality triangles
concentrated at centers of faces of the
original icosahedron.  On the other hand, the SAC configuration, which provides a lower mesh
ratio (higher quality point configuration) and tends to have a more even distribution of edge ratios
throughout its triangulation.  

Fig.~\ref{fig2c} shows the first application of the general SAC method presented here (e.g.,
with $K>1$).  For the same $N=252$, this SAC approach provides a mesh ratio that is a significant
improvement over Figs.~\ref{fig2a} and~\ref{fig2b}, and a triangulation that shows a more even
distribution of edge ratios.  

For further evaluations, consider the configurations in
Figs.~\ref{fig3a}--\ref{fig3f} with $N$ chosen larger (in the range $7000$--$8000$).  It can be
seen comparing
Fig.~\ref{fig3a} and Fig.~\ref{fig3b} that the EQA approach tends to have higher-quality
triangles, yet overall the SAC method achieves a lower mesh ratio (recall, the mesh ratio is a
global quantity) for the same $N$.  Fig.~\ref{fig3c} shows that for a slightly higher value of $N$ this SAC method is
able to achieve a lower mesh ratio with well-chosen input parameters (note: mesh ratios
tend to increase with $N$).  Furthermore, for $N=7682$,
Figs.~\ref{fig3d} and~\ref{fig3e} demonstrate alternative, improved, choices of integer-pair
sequences.   Finally, Fig.~\ref{fig3f} shows a slight variation and improvement to that in
Fig.~\ref{fig3c}.  It is clear from Figs.~\ref{fig3c}--\ref{fig3f} that a lower mesh ratio is
achieved through a more uniform distribution of triangle qualities, rather than the appearance
of high-quality triangles in local region.

\begin{figure*}[h]
      \centering
      \includegraphics[width=\textwidth,clip,trim=2cm 0cm 2cm 0cm]{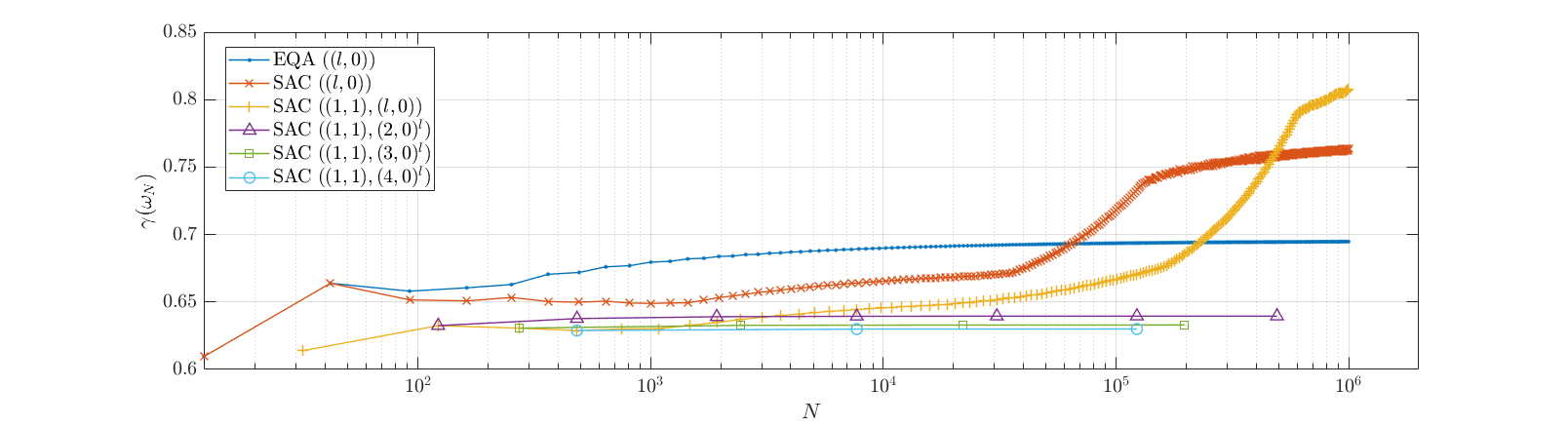}
      \caption{Mesh ratios for sequences of $N$-point spherical configuration with $N<10^6$ (with
      integer $l>0$ increasing within each sequence).}
      \label{fig4}
\end{figure*}

For the final evaluation of this approach in this study, we investigate the behavior of 
\emph{sequences} of $N$-point spherical configurations that may be generated with this SAC approach.
Shown in Fig.~\ref{fig4} are calculated mesh ratios for $N$-point spherical configurations with $N$ limited to
$10^6$.  The sequences of spherical configurations featured here are given by parameterised
integer-sequences with one integer
parameter $l>0$.  The first sequence, labeled ``EQA $((l,0))$'', is a sequence from the Icosahedral
Equal-Area method featured in~\cite{michaels2017equidistributed} with the lowest reported
limiting mesh
ratio for any sequence in the literature (to the author's knowledge).  The following sequences
use the general SAC approach.  It can
be seen that for $N<60,000$, the SAC sequence ``$((l,0))$'' outperforms the EQA approach in terms of calculated
mesh ratios.  Although, a better choice for $N<200,000$, is the SAC sequence ``$((1,1),(l,0))$''.
It is important to note that limiting mesh ratios for those SAC sequences do not appear to be 
bounded for large $N$, so they lack the property of
\emph{quasi-uniformity}~\cite{michaels2017equidistributed}.

The final three recursive sequences (EQA $((1,1),(2,0)^l)$, etc.) show the lowest mesh ratios of
all, but are sparser in terms of $N$ values.  Mesh ratios for those recursive
sequences do appear to be bounded as $N$ increases, which hints at the property of
quasi-uniformity.  In particular, the sequence given by $((1,1),(4,0)^l$ has a
mesh ratio that appears to be bounded by 0.630, which is significantly lower than the best EQA
sequence from~\cite{michaels2017equidistributed}.

\section{Conclusions}
A method for generating $N$-point spherical configurations was presented in this short article.  The
method features Caspar-Klug triangular point grids extended to non-equilateral spherical triangles
using spherical area coordinates (SAC).  This SAC method can be applied recursively and is
parameterised by a sequence of integer pairs.  The method was evaluated numerically and
demonstrated the ability to return $N$-point spherical configurations with mesh ratios lower
than previously reported in the literature.  

The method presented herein is rather general, but may require careful selection of input parameters
(integer-pair sequences) for good results.  Some favorable examples are given here, but it is not
always clear which parameters are optimal for a given $N$.  Future work could investigate more
choices of integer-pair sequences with this approach, as well as evaluate such configurations
with other measures (e.g., potential energies).  Only icosahedral configurations were
investigated for brevity, but the method works well with octahedral point sets.  Comparsions
with existing octahedral configurations is left for future work.  

Python code to generate spherical
configurations with this method is provided at~\url{https://github.com/bsxfun/sac-method}.


\bibliographystyle{elsarticle-num} 
\bibliography{refs.bib}

\begin{thebibliography}{10}
\expandafter\ifx\csname url\endcsname\relax
  \def\url#1{\texttt{#1}}\fi
\expandafter\ifx\csname urlprefix\endcsname\relax\def\urlprefix{URL }\fi
\expandafter\ifx\csname href\endcsname\relax
  \def\href#1#2{#2} \def\path#1{#1}\fi

\bibitem{saff1997distributing}
E.~B. Saff, A.~B. Kuijlaars, Distributing many points on a sphere, The
  Mathematical Intelligencer 19~(1) (1997) 5--11.

\bibitem{koay2011analytically}
C.~G. Koay, Analytically exact spiral scheme for generating uniformly
  distributed points on the unit sphere, Journal of Computational Wcience 2~(1)
  (2011) 88--91.

\bibitem{womersley2018efficient}
R.~S. Womersley, Efficient spherical designs with good geometric properties,
  in: Contemporary computational mathematics-A celebration of the 80th birthday
  of Ian Sloan, Springer, 2018, pp. 1243--1285.

\bibitem{caspar1962physical}
D.~L. Caspar, A.~Klug, Physical principles in the construction of regular
  viruses, in: Cold Spring Harbor symposia on quantitative biology, Vol.~27,
  Cold Spring Harbor Laboratory Press, 1962, pp. 1--24.

\bibitem{holhocs2014octahedral}
A.~Holho{\c{s}}, D.~Ro{\c{s}}ca, An octahedral equal area partition of the
  sphere and near optimal configurations of points, Computers \& Mathematics
  with Applications 67~(5) (2014) 1092--1107.

\bibitem{michaels2017equidistributed}
T.~Michaels, Equidistributed icosahedral configurations on the sphere,
  Computers \& Mathematics with Applications 74~(4) (2017) 605--612.

\bibitem{hardin2016comparison}
D.~P. Hardin, T.~Michaels, E.~B. Saff, A comparison of popular point
  configurations on $\mathbb{S}^2$, arXiv preprint arXiv:1607.04590 (2016).

\bibitem{bondarenko2014mesh}
A.~Bondarenko, D.~P. Hardin, E.~B. Saff, Mesh ratios for best-packing and
  limits of minimal energy configurations, Acta Mathematica Hungarica 142~(1)
  (2014) 118--131.

\bibitem{bauer2000distribution}
R.~Bauer, Distribution of points on a sphere with application to star catalogs,
  Journal of Guidance, Control, and Dynamics 23~(1) (2000) 130--137.

\bibitem{lei2020new}
K.~Lei, D.~Qi, X.~Tian, A new coordinate system for constructing spherical grid
  systems, Applied Sciences 10~(2) (2020) 655.

\bibitem{pebay2003analysis}
P.~P{\'e}bay, T.~Baker, Analysis of triangle quality measures, Mathematics of
  computation 72~(244) (2003) 1817--1839.

\end{thebibliography}


%
%
%
\end{document}